\documentclass[twocolumn,superscriptaddress,showpacs,preprintnumbers,amsmath,amssymb,aps,pre]{revtex4}
\usepackage{graphicx}
\usepackage{dcolumn}
\usepackage{bm}
\begin{document}
\title{Similarity of fluctuations in correlated systems: The case of seismicity\footnote{Published in Physical Review E {\bf72}, 041103 (2005).}}
\author{P. A. Varotsos}
\email{pvaro@otenet.gr}
\affiliation{Solid State Section, Physics Department, University of Athens, Panepistimiopolis, Zografos 157 84, Athens, Greece}
\affiliation{Solid Earth Physics Institute, Physics Department, University of Athens, Panepistimiopolis, Zografos 157 84, Athens, Greece}
\author{N. V. Sarlis}
\affiliation{Solid State Section, Physics Department, University of Athens, Panepistimiopolis, Zografos 157 84, Athens, Greece}
\author{H. K. Tanaka}
\affiliation{Earthquake Prediction Research Center, Tokai University 3-20-1, Shimizu-Orido, Shizuoka 424-8610, Japan}
\author{E. S. Skordas}
\affiliation{Solid Earth Physics Institute, Physics Department, University of Athens, Panepistimiopolis, Zografos 157 84, Athens, Greece}
\begin{abstract}
We report a similarity of fluctuations in equilibrium critical phenomena and non-equilibrium
 systems, which is based on the concept of natural time.
The world-wide seismicity as well as that of San Andreas fault
system and Japan are analyzed. An order parameter is chosen and its fluctuations relative to the standard
deviation of the distribution are studied. We find that the scaled distributions fall on the same curve,
which interestingly exhibits, over four orders of magnitude, features similar to those in several
equilibrium critical phenomena ( e.g., 2D Ising model) as well as in non-equilibrium systems
(e.g., 3D turbulent flow).
\end{abstract}
\maketitle

\section{introduction}
Recently, a great interest has been focused on the fluctuations of correlated systems in
general and of critical systems in particular\cite{BRA98,BRA00,BRA01,ZHE01,BRA00R1,BRA00R2,BRA00R3,ZHE03,CLU04}.
 Bramwell, Holdsworth and Pinton (BHP)\cite{BRA98}, in an experiment of
a closed turbulent flow, found that the ({\em normalized})
probability distribution function (PDF) of
the power fluctuations has the same functional form as that of the magnetization ($M$) of the
finite-size 2D (two-dimensional) XY equilibrium model in the critical region below the
Kosterlitz-Thouless transition temperature (Magnetic ordering is then described by the {\em order parameter} $M$).
 The normalized PDF, denoted by $P(m)$, is defined by introducing the
reduced magnetization\cite{BRA98} $m=(M-\langle M\rangle )/\sigma$, where $\langle M\rangle$
 denotes the mean and  $\sigma$ the standard
deviation. For both systems, BHP found that while the high end ($m>0$) of the distribution
has\cite{BRA98} a Gaussian shape the asymptote of which was later clarified\cite{BRA01} to have 
a double exponential 
form, a distinctive exponential tail appears towards the low end ($m<0$) of
the distribution. The latter tail, which will be hereafter simply called, for the sake of convenience,
``exponential tail'', provides the main
region of interest\cite{BRA98}, since it shows that the probability for a rare fluctuation, e.g., of
greater than six standard deviations from the mean, is almost five orders of magnitude higher than in
the Gaussian case.  Subsequent independent simulations
\cite{BRA00,BRA01,ZHE01,ZHE03,CLU04} showed that a variety of highly
correlated (non equilibrium as well as equilibrium)
systems, under certain conditions, exhibit approximately the ``exponential tail''.

Earthquakes do exhibit complex correlations in space, time and magnitude, e.g. \cite{BAK02,COR04,ABE04}.
It has been
repeatedly proposed (see Ref.\cite{SOR04} and references therein) that the occurrence of earthquakes (cf. mainshocks) can be considered as
a critical point (second-order phase change), but alternative models based on first-order phase transitions have 
been also forwarded which are probably more applicable, see Ref.\cite{RUN03} and references therein. 
(Such a diversity also exists for the brittle rupture which is a phenomenon closely 
related to earthquakes. Buchel and Sethna\cite{BUC97} have associated brittle rupture with 
a first-order transition and a similar view has been also expressed in Refs.\cite{ZAP97,KUN99}. On the 
other hand, Gluzman and Sornette\cite{GLU01} later suggested that it is 
analogous to a critical point phenomenon.) Both approaches lead to scaling laws or 
power-law distributions for the dynamical variables (second-order transition demonstrate 
scaling near a critical point, whereas first-order transitions demonstrate scaling when 
the range of interactions is large (mean-field condition), as is the case with elastic 
interactions\cite{RUN03}).  However, the question  on whether
earthquakes exhibit an  ``exponential tail'', has not yet been clarified. This might be due to the major difficulty
of choosing an order parameter in the case of earthquakes(EQs). Following the wording of Ref.\cite{SET92}, we note that
in general such a choice is an art, since usually it's a new phase which we do not understand yet, and guessing the
order parameter is a piece of figuring out what's going on. The scope of the present paper is twofold: to propose
an order parameter for the case of EQs and then examine whether an ``exponential tail'' appears.
We find that  our scope is achieved only  {\em if} we analyze the series of earthquakes in the
natural time-domain\cite{VAR01,VAR02,VAR02A,VAR03A,VAR03B,VAR04,VAR05,VAR05B}.

In order to serve the aforementioned scope, the present paper is organized as follows:
 In Section II, we explain how the power  spectrum of the seismicity in natural time can be obtained.
An order parameter for EQs is proposed in Section III. In the light of this proposal,
and without using any adjustable parameter, we show in Section IV that
the normalized distribution of the long term seismicity for different seismic areas fall on a universal curve. It consists of two segments the one of which exhibits the ``exponential tail''. Interestingly, a further investigation of the latter segment in Section V, reveals that it is similar to that observed in several equilibrium critical phenomena (e.g., 2D Ising, 3D Ising) and in non-equilibrium systems (e.g., 3D turbulent flow).  A brief discussion follows in Section VI, while Section VII summarizes our main conclusions. Two Appendices provide clarifications
on some points discussed in the main text.

\section{the seismicity  in natural time}

\begin{figure}
\includegraphics{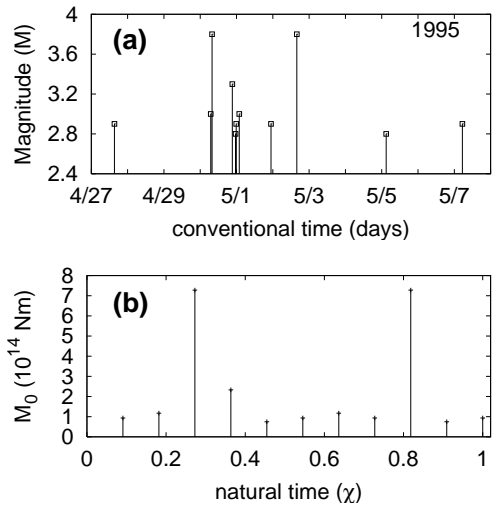}
\caption{\label{fg1} How a series of seismic events in conventional time (a) is read in the natural time (b).
This example refers to the first 11 small earthquakes (cf. the month/date is marked on the horizontal axis in (a)) that occurred after the SES activity recorded on April 18,1995 and preceded the mainshock (M=6.6) of May 13,1995.}
\end{figure}

In a time series consisting of $N$ events, the {\em natural time} $\chi_k = k/N$ serves as an index\cite{VAR01,VAR02}
 for the occurrence of the $k$-th event.  It is, therefore, smaller than, or equal to, unity.
For the analysis of seismicity, the evolution of the pair ($\chi_k, E_k$)
is considered\cite{VAR01,VAR02A,VARBOOK,TAN04}, where $E_k$ denotes the seismic energy released during the
$k$-th event see Fig.\ref{fg1} (cf. This energy -which is itself proportional to the seismic moment $M_0$
and hence we can use in the vertical axis of Fig.\ref{fg1}(b) either $E_k$ or $(M_0)_k$ is
related\cite{EPAPS} to the magnitude M through $E\propto  10^{c {\rm M}}$, where $c$ is a constant around 1.5).
The following continuous function $F(\omega )$   was introduced\cite{VAR01,VAR02,VAR02A}:
$F(\omega)=\sum_{k=1}^{N} E_{k}
\exp \left( i \omega \frac{k}{N} \right)$
where $\omega =2 \pi \phi$, and $\phi$ stands for the
{\it natural frequency}. We normalize $F( \omega )$
by dividing it by $F(0)$,
\begin{equation}
\Phi(\omega)=\frac{\sum_{k=1}^{N} E_{k}
\exp \left( i \omega \frac{k}{N} \right)}{\sum_{n=1}^{N} E_{n}}=
\sum_{k=1}^{N} p_k \exp \left( i \omega \frac{k}{N} \right)
\label{equ1}
\end{equation}
where $p_k=E_{k}/\sum_{n=1}^{N}E_{n}$. A kind of normalized
power  spectrum $\Pi(\omega)$ can now be defined:
$\Pi(\omega)=\left| \Phi(\omega) \right|^2$.

For a Seismic Electric Signals (SES) activity, which is a sequence of low frequency 
($\leq 1$ Hz) electric pulses emitted when the stress in the 
focal area approaches\cite{VAR86,VAR03C} a {\em critical } value, 
we have shown (for details see Ref.\cite{VAR01}, see also \cite{VAR02}) that  the following relation holds\cite{VAR01,VAR02,VAR02A}
\begin{equation}
\Pi ( \omega ) = \frac{18}{5 \omega^2}
-\frac{6 \cos \omega}{5 \omega^2}
-\frac{12 \sin \omega}{5 \omega^3}.
\label{fasma}
\end{equation}

We focus on the properties of $\Pi(\omega)$ or $\Pi(\phi)$ for natural frequencies $\phi$ less than 0.5, since in
this range of $\phi$, $\Pi(\omega)$  or $\Pi(\phi)$ reduces\cite{VAR01,VAR02,VAR02A,VARBOOK}
 to a {\em characteristic function} for the
probability distribution $p_k$  in the context of probability theory. According to the probability
theory, the moments of a distribution and hence the distribution itself can be approximately determined once
the behavior of the characteristic function of the distribution is known around zero. For
$\omega \rightarrow 0$, Eq.(\ref{fasma}) leads to\cite{VAR01,VAR02,VARBOOK}:
\begin{equation}
\label{eq4}
\Pi (\omega )\approx 1-0.07 \omega^2
\end{equation}
which reflects (see Appendix \ref{apb}) that the variance of $\chi$ is given by:
\begin{equation}
\label{eq5}
\kappa_1=\langle \chi^2 \rangle -\langle \chi \rangle ^2=0.07.
\end{equation}
In Section IV, we will investigate whether  Eq.(\ref{fasma}) holds for EQs.

\section{the order parameter proposed}
\begin{figure}
\includegraphics{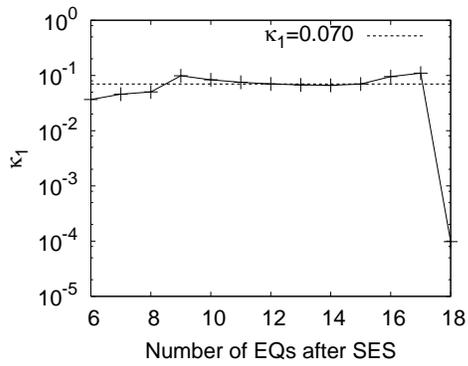}
\caption{\label{fg3} How the variance $\kappa_1$ evolves event by event during the following
period: from the detection\cite{VAR03C} of the SES activity on April 18, 1995 until
the occurrence of the M=6.6 mainshock(labelled 18) on May 13, 1995.
 All the EQs used in the calculation are tabulated in
Ref.\cite{EPAPS}. (cf. the first 11 EQs -out of 18- are those depicted in Fig.\ref{fg1}(a)). }
\end{figure}

We now proceed to choose the order parameter, assuming that a mainshock may be
considered as the new phase.
We take advantage of the experimental fact\cite{VAR03C} that several hours to a few
months before a mainshock an SES activity is recorded, 
and  focus our attention on the evolution of the seismicity (in the candidate area) during
the period from the SES detection until the mainshock.
 If we set the natural time for the seismicity zero at the initiation
of the concerned SES activity, we form time series of seismic events in natural time (see Fig.\ref{fg1}) for
various time windows as the number $N$ of consecutive (small) EQs  increases. When computing $\Pi (\phi )$
(as well as $\kappa_1$, see below) for each of the time windows, we find that,
 in the range $0<\phi \leq 0.5$,
it approaches, as $N$ increases from 6 to some value less than (or equal to) 40, to that given by Eq.(\ref{fasma})
(or the $\kappa_1$-value becomes equal to 0.07, see Eq.(\ref{eq5})). The coincidence occurs {\em only} a few
hours to a few days before the mainshock. (In simple words, before
a mainshock  a sequence of earthquakes occurs, which obeys Eq.(\ref{fasma}) and this process will be called
{\em single correlated process}.)  When the mainshock occurs (the new phase), $\Pi (\phi )$
abruptly increases to approximately unity (for details see Ref.\cite{EPAPS}) and $\kappa_1$ becomes almost zero.
 This can be visualized in the
example depicted in Fig.\ref{fg3}, where we plot the $\kappa_1$-value versus the number of EQs after the SES
detection on April 18, 1995 (see Refs.\cite{VAR02,VAR03C}) until the occurrence of the M=6.6
mainshock  on May 13, 1995 at $40.2^o N, 21.7^oE$. This figure shows that the $\kappa_1$-value becomes $\kappa_1\approx 0.07$ after the 11th EQ (see also Ref.\cite{EPAPS}), while upon the mainshock the $\kappa_1$-value
{\em abruptly} decreases to $\kappa_1 \approx 9 \times 10^{-5}$.
 Such a behavior has been verified\cite{VAR01,VAR02A} for
several major EQs and points to the conclusion that $\Pi (\phi )$ for small $\phi$, or $\kappa_1$, could be considered
as an order parameter.

\section{universal curve for seismicity}

The properties of the power spectrum for  the long term seismicities in natural time can be 
studied by means of the following
procedure: First, calculation of   $\Pi(\phi)$ was made for an event taking time
windows from 6 to 40 consecutive events 
(for the reasons explained in Section III; the choice of the precise value of the upper limit, up to 100 or so, is not found decisive\cite{VAR01,VAR02A,TAN04}).  And second,
this process was performed for all the events by scanning the whole catalogue. The following
data from two different areas, i.e., San Andreas fault system and Japan, have been analyzed:
First, the EQs that occurred during the period 1973-2003 within the area $N_{32}^{37} W_{114}^{122}$
using the Southern California Earthquake catalogue (hereafter called SCEC). Second, the
EQs within $N_{25}^{46} E_{125}^{146}$ for the period 1967-2003 using the Japan Meteorological
Agency catalogue (hereafter simply called ``Japan''). The thresholds M$\geq$2.0 and M$\geq$3.5 have
been considered for SCEC and Japan, respectively, for the sake of data completeness\cite{EPAPS}.
 By plotting for a given value of $\phi$ the observed probability
$P [\Pi(\phi)]$ versus $\Pi(\phi )$ (two such examples are given in Figs.\ref{fg2}(a) and \ref{fg2}(b) for
$\phi=0.05$ and $\phi=0.4$, respectively), we find that a local
maximum occurs at a value of $\Pi(\phi)$ hereafter called $\Pi_p(\phi)$ (see also Appendix \ref{apc}).
 This lies very close (see Fig.\ref{fg2}(c))
to the value $\Pi_{th}(\phi )$ obtained theoretically, i.e., estimated from Eq.(\ref{fasma}).
The validity of Eq.(\ref{fasma}) for various $\phi$-values, in the range $0<\phi \leq 0.5$,
 can be now  visualized in Fig.\ref{fg2}(d), where we see that  $\Pi_p(\phi)$-values versus
 $\phi$ for both SCEC and Japan  do not differ
by more than 1\% from the $\Pi_{th}(\phi )$-values (cf. this difference is more or less comparable to the estimation error of $\Pi_p(\phi)$, for details see Appendix \ref{apc}).

We now plot, in Fig.\ref{fg4}, the quantity $\sigma P(X)$ versus
 $(X-\langle X\rangle)/\sigma$ where $X$ stands for
$\Pi(\phi)$  and $\langle \Pi(\phi) \rangle$ and $\sigma$ refer to the mean value and the standard deviation of
$\Pi(\phi)$   (recall that the
calculations should be done for small $\phi$-values,
e.g., $\phi$=0.05, since we assume here $\phi \rightarrow 0$, for the
reasons explained in Section II). One could alternatively plot $\sigma_{\kappa_1}P(\kappa_1)$ versus
 $(\langle \kappa_1 \rangle -\kappa_1)/\sigma_{\kappa_1}$,
where $\langle \kappa_1 \rangle$ and $\sigma_{\kappa_1}$ now refer to the mean value and
the standard deviation of $\kappa_1$. The results
in Fig. \ref{fg4}, for both areas, fall on the {\em same} curve. This log-linear plot clearly consists of two
segments: The segment to the left shows a decrease of $P(X)$ almost by five orders of
magnitude, while the upper right segment has an almost constant $P(X)$ (Obviously, the latter segment
deviates from the general behavior of the BHP distribution -as it was summarized in Section I-
but  from thereon we put emphasis on the left segment since our main interest here concerns the ``exponential tail'').
  The feature of this
plot is strikingly reminiscent of the one obtained by Bak et al.\cite{BAK02} (see their Fig. 4) on
different grounds, using EQs in California only. More precisely, they measured $P_{S,l}(T)$,
 the distribution of waiting times $T$, between EQs occurring within range $l$ whose
magnitudes are greater than $M\equiv \log S$. They then plotted $T^\alpha P_{S,l}(T)$ versus
$TS^{-b}l^d$  and found
that, for a {\em suitable choice} of the exponent $\alpha$ (i.e., $\alpha=1$),
the Gutenberg-Richter law exponent $b$
(i.e., $b$=1) and the spatial dimension $d$ (i.e., fractal dimension $d=1.2$)
 all the data collapse onto
a {\em single} curve which is similar to that of Fig.\ref{fg4}. Recall, however,
that Fig.\ref{fg4}
  was obtained here
without considering at all the waiting time distribution and without the suitable choice of any
parameter. After a further inspection of Fig.\ref{fg4}, the following points have been clarified:

First, the rapidly decaying part (i.e., the left segment), which is consistent with
an almost exponential decaying function over almost four orders of magnitude, remains
practically unchanged, upon {\em randomizing} the data (``shuffling''\cite{VAR04}). (Some  changes
do occur in the right part, associated with aftershocks, see also below.) This can be seen in the
inset of Fig. \ref{fg4}, where for the sake of clarity only the results from the data of Japan (the
original as well as the ``shuffled'' ones) are depicted.

Second, the feature of the plot of Fig. \ref{fg4} is not altered upon changing either the seismic
region or the time-period (provided that the latter does not include aftershocks {\em only}, see
below). As an example,  Fig.\ref{fg5}(b) shows that three different regions A, B, C in Japan (depicted in Fig.\ref{fg5}(a)),
as well as the whole Japan, result in almost identical plots.

Third, the ``upturn branch'' in the upper right part of Fig. \ref{fg4} arises from the presence of
aftershocks. It disappears (see the
crosses in Fig. \ref{fg6}) when, in Japan, for example, we delete the EQs with M$\leq$5.7 (and
hence drastically reduce the number of aftershocks), but it does not, when deleting
EQs with smaller threshold, i.e., M$<$4.0; the latter can be also visualized in the SCEC
example of Fig. \ref{fg6}, where we give the results for M$\geq$4.0 (cf. this threshold still allows the
presence of a reasonable number of aftershocks).

Fourth, if we consider the relevant results for the worldwide seismicity (WWS) by
taking a large magnitude threshold, i.e., M$>$5.7 (so that for the data to be complete\cite{EPAPS}),
 we find (see Fig.\ref{fg6} that will be further discussed below) that they fall 
 onto the {\em same} curve with the results of
both Japan and SCEC.

\section{does a universal behaviour exist for diverse systems?}
We now compare in Fig.\ref{fg6} the aforementioned results of seismicity  with those obtained in some equilibrium critical systems (e.g., see Ref.\cite{ZHE03}). We first recall that the PDF in the {\em critical} regime depends on $K = 1/T$ and the length $L$
through a scaling variable $s\equiv L^{1/\nu} (K-K_c)/K_c$, where $K_c = 1/T_c$ and $T_c$ denotes the critical
temperature (the quantity $s^\nu$ provides the ratio of the lattice size and the {\em correlation length} at
$K$). In Fig. \ref{fg6}, we include numerical results of the 2D Ising model for $s=8.72 (L=128,
K=0.4707)$ and $s=17.44 (L=256, K=0.4707)$. Here, $X$ stands for $M$. These $s$-values were
intentionally selected, because\cite{ZHE03} for $s\geq 8.72$ for the 2D Ising model, the $P(m, s)$'s of a
number of critical models (i.e., 2D XY, 2D Ising, 3D Ising, 2D three-state Potts) share the
{\em same} form (up to a constant factor of $s$), which interestingly exhibits an exponential-like left-tail ($m<0$). 
An inspection of Fig.\ref{fg6} shows that our 2D Ising results almost coincide (cf. this can be safely checked only for the left segment, i.e., $m<0$)
with those of seismicity, i.e., Japan, SCEC and WWS (cf. Some disparity which appears in the upper right
part of SCEC only, might be attributed to the selection of the magnitude threshold for seismicity, recall the third
point mentioned in Section IV). This coincidence (which seems to be
better for $s=17.44$) reveals that the seismicity, irrespective of the seismic area we consider,
exhibits -over four orders of magnitude- fluctuations of the order parameter similar to those in
several critical systems as well as in 3D turbulent flow.

\section{Discussion}
 
It is of interest to see how the scaled distributions look like in
the frame of the present analysis, if one generates surrogate data
either by means of a simple Poisson model or by the
Gutenberg-Richter law and compare the results to those deduced
from actual seismicity data.

In Fig.\ref{fgPoi}, we present the linear-linear plot (Fig. \ref{fgPoi}(a)) as well as
the log-linear plot (Fig. \ref{fgPoi}(b)) of $\sigma P(X)$ versus $(X- \langle X
\rangle)/ \sigma$ where $X$ stands for $\Pi(\phi)$ for $\phi
\approx 0$, for surrogate data of EQs for which their $(M_o)_{k}$
obey a simple Poisson rule for
various mean values $\lambda$ lying between 5 and 200.
In the same figure we insert the results for Japan (M $\geq 3.5$)
and SCEC (M $\geq 2.0$) already discussed in Fig.\ref{fg4}. Although we
find that upon decreasing $\lambda$ the surrogate data move closer
to the real data, however a satisfactory agreement between them
cannot be supported.

In Fig. \ref{fgGuRi} we repeat the procedure followed in Fig. \ref{fgPoi}, but now the
surrogate data are produced on the basis of the Gutenberg-Richter
law, i.e., that the (cumulative) number of EQs with magnitude
greater than M (occurring in a specified region and time) is
given by 
\begin{equation}
\label{aguri}
N(> {\rm M}) \approx 10^{-b{\rm M}}.
\end{equation}
 It is currently
considered\cite{RUN03} that $b$ is generally a constant varying
only slightly from region to region being approximately in the
range $0.8 \leq b \leq 1.2$. For Japan and SCEC we find on the basis of 
Eq.(\ref{aguri}) $b\approx 1.05$. Note that in Fig.\ref{fgGuRi}, surrogate data
are intentionally produced for a variety of $b$ values in the
range $b=0.5$ to 2.0. An inspection of this figure leads to the
following conclusions: First, the curves of the surrogate data
marked with $b=0.5$ to $b=0.9$ significantly differ from that of
the real data. Second, for $b$-values larger than 1 and smaller than 1.4,
  the curves of the surrogate data have a general
feature more or less similar to the curve of the real data.
However, none of these $b$-values in the surrogate data can lead to a curve
coinciding to the one obtained from the real data. 

In other words,
the scaled distribution, deduced within the frame of the present
analysis, reveals for the real data an extra complexity  when compared to
the surrogate data even if the latter are produced with 
$b$-values comparable to the experimental ones.

\section{conclusions}
The main conclusions could be summarized as follows:
 
 (1)The analysis of the seismicity in the natural time-domain reveals that
 $\Pi(\phi)$ (for small  $\phi$) or $\kappa_1$, may be considered  as an order parameter.

(2)If we study the order parameter fluctuations relative to the standard deviation of its distribution, 
the following two facts emerge (without making use of {\em any} adjustable 
parameter):

First, the scaled distributions of different seismic areas (as well as that of the world wide seismicity)
fall on the {\em same} curve ({\em universal}).

Second, this curve exhibits  an  ``exponential tail'' form similar to that
 observed in certain non-equilibrium
systems (e.g., 3D turbulent flow) as well as in several(e.g., 2D Ising, 3D Ising, 2D XY)
 equilibrium critical phenomena.

\appendix
\section{derivation of equation (\ref{eq5})}
\label{apb}
The Taylor expansion, around $\omega$=0, of the relation $\Pi(\omega)=\left| \Phi(\omega) \right|^2$  using
Eq(1) reveals that \cite{VAR01}
\begin{equation}
\Pi(\omega)=1-\kappa_1\omega^2+\kappa_2\omega^4+\kappa_3\omega^6+\kappa_4\omega^8+ \ldots
\end{equation}
where
\begin{equation}
 \kappa_1=- \left. \frac{1}{2}\frac{d^2 \Pi(\omega)}{d\omega^2}\right|_{\omega=0}.
\end{equation}
We now consider
\begin{equation}
\frac{d^2\Pi(\omega)}{d\omega^2}={ \Phi}^{*}(\omega) \frac{d^2\Phi(\omega )}{d\omega^2}+
\Phi(\omega)\frac{d^2\Phi ^*(\omega)}{d\omega^2}+2 \frac{d\Phi(\omega)}{d\omega}
\frac{d\Phi ^*(\omega)}{d\omega}
\end{equation}
and taking into account that $\Phi(\omega)\equiv \sum_{k} p_k \exp ( i\omega \chi_k )$, with $\Phi(0)=1$, we find:
\begin{eqnarray}
&\kappa_1&=-\frac{1}{2} \left[ -\sum_{k}p_k \chi_k^2-\sum_{k}p_k \chi_k^2+2 \left( \sum_{k}p_k \chi_k \right)^2  \right] \nonumber \\
&=&\langle \chi^2 \rangle - \langle \chi \rangle^2,
\end{eqnarray}
where $\langle \chi^n \rangle = \sum_{k}p_k \chi_k^n$.

Expanding Eq.(\ref{fasma}), around $\omega=0$, we have:
\begin{equation}
\Pi(\omega)=1-0.07 \omega^2+\ldots
\end{equation}
and hence:
\begin{equation}
\kappa_1=\langle \chi^2\rangle - \langle \chi\rangle^2=0.07
\end{equation}
which is just Eq.(\ref{eq5}).

\section{the procedure to determine the maximum in $P[\Pi(\phi)]$ versus $\Pi(\phi)$}
\label{apc}
\begin{figure}
\includegraphics{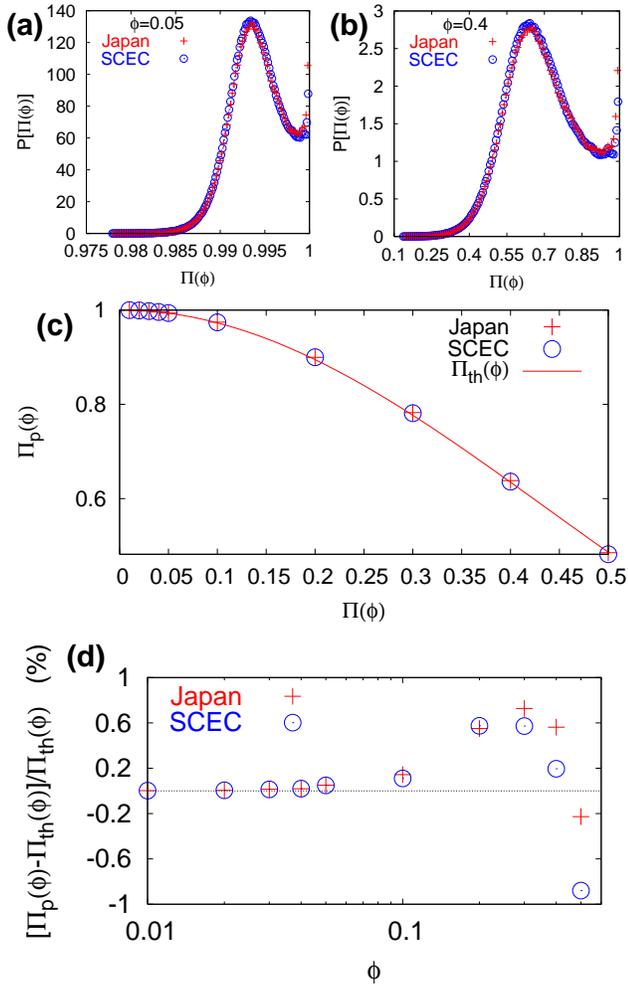}
\caption{\label{fg2} (Color online) Validity of Eq.(\ref{fasma}) for SCEC and Japan.
We first determine (see also Appendix B) for SCEC(circles) and Japan(crosses), for each $\phi$-value, the value $\Pi_p(\phi)$
at which $P[\Pi(\phi)]$ maximizes. Two such examples are shown in (a) and (b) for
$\phi=0.05$ and $\phi=0.4$, respectively.
In (c), we plot the
resulting $\Pi_p(\phi)$ values versus $\phi$ for SCEC(circles) and Japan(crosses); the solid line corresponds to Eq.(\ref{fasma}). The percentage difference
between $\Pi_p(\phi)$ and $\Pi_{th} (\phi )$ (obtained from Eq.(\ref{fasma})) is shown in (d).}
\end{figure}

\begin{figure}
\includegraphics{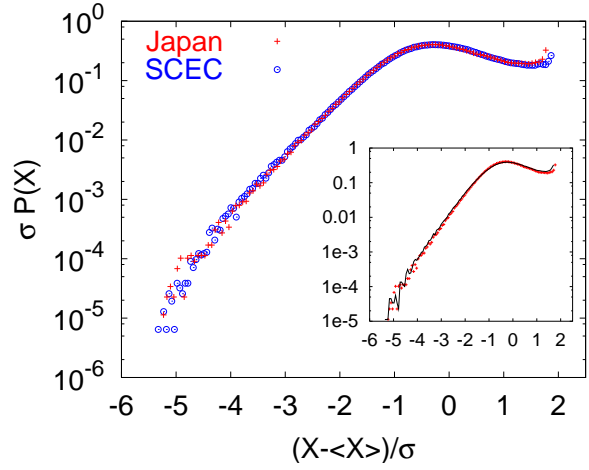}
\caption{\label{fg4} (Color online)  Universality of the probability density function of
$\Pi(\phi )$ for EQs in the natural
time-domain. The log-linear plot of $\sigma P(X)$ versus $(X-\langle X \rangle )/\sigma$,
where $X$ stands for $\Pi(\phi )$ for $\phi \approx 0$.
 Crosses  and circles  correspond to Japan (M$\geq$3.5) and SCEC
(M$\geq$2.0), respectively. The inset depicts the corresponding results for the
``shuffled'' data
(black curve) and the original data (red crosses) in Japan.  The same graph is obtained for
 three different regions in Japan (see Fig.\ref{fg5}).}
\end{figure}

\begin{figure}
\includegraphics{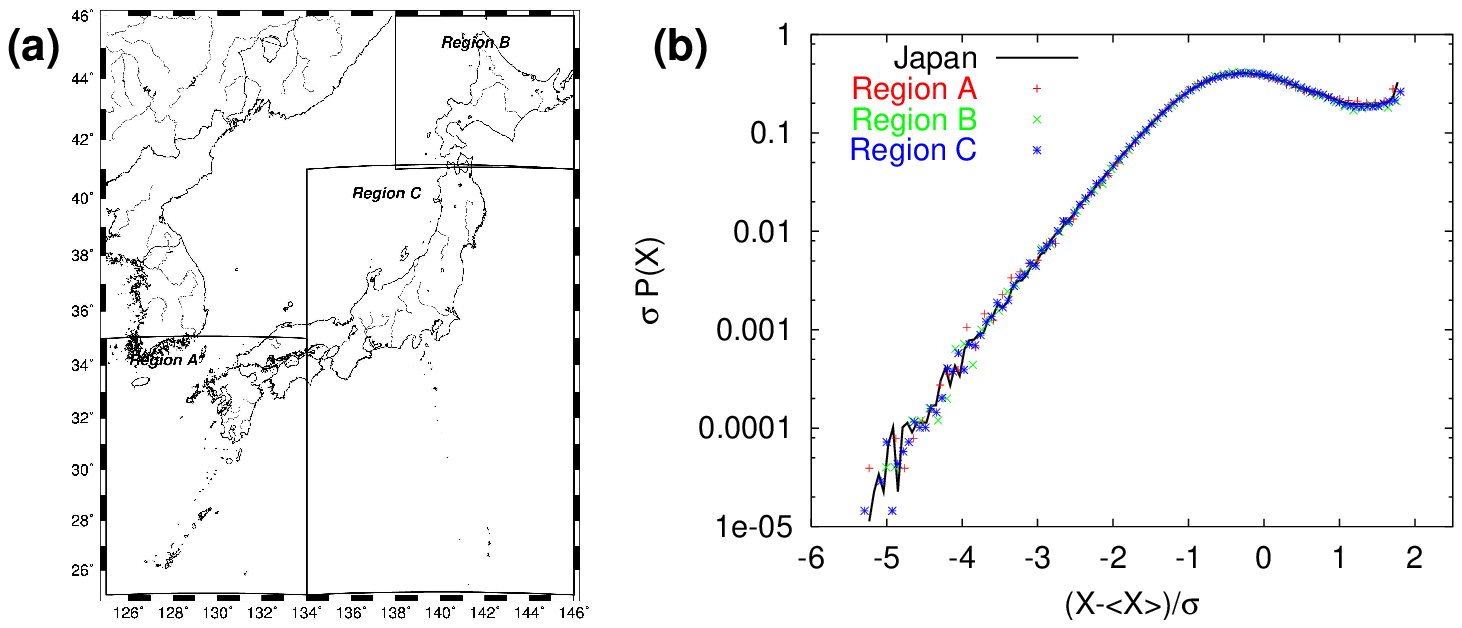}
\caption{\label{fg5} (Color online) The same as Fig. \ref{fg4}, but
for the regions A (red), B (green) and C (blue) in Japan (b). A
map of these regions is shown in (a).}
\end{figure}

\begin{figure}
\includegraphics{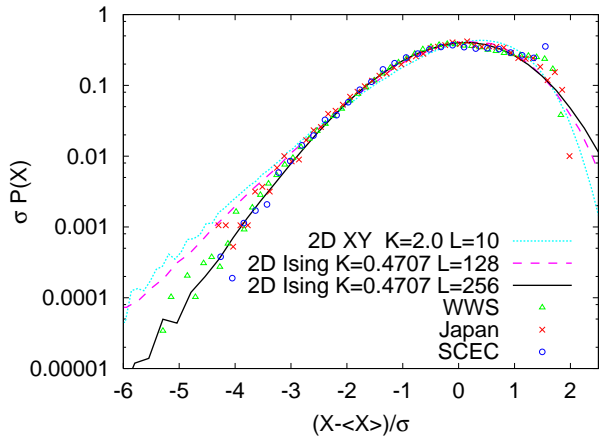}
\caption{\label{fg6}(Color online) The common feature of fluctuations in different
 correlated systems. The log-linear plot
of $\sigma P(X)$ versus $(X-\langle X\rangle)/\sigma$ for
WWS(triangles), Japan(crosses) and SCEC(circles). The magnitude
threshold M$>$5.7 for WWS and Japan (while M$\geq$4.0 for SCEC)
was used see the text.  Furthermore, the dotted curve shows the
results obtained for the 2D XY model (with\cite{STR94} inverse
Kosterlitz-Thouless transition temperature $K_{KT}\approx1.2$) ($X=\sqrt{M_x^2+M_y^2}$), $K=2.0$ for $L=10
(N=100)$ which has been shown\cite{BRA98} to describe the
experimental results for 3D turbulent flow. The results of the  2D Ising
model $K=0.4707$ (while $K_c\approx0.4407$), either for
$L=128$(dashed) or $L=256$(solid line), are also plotted.}
\end{figure}

\begin{figure}
\includegraphics{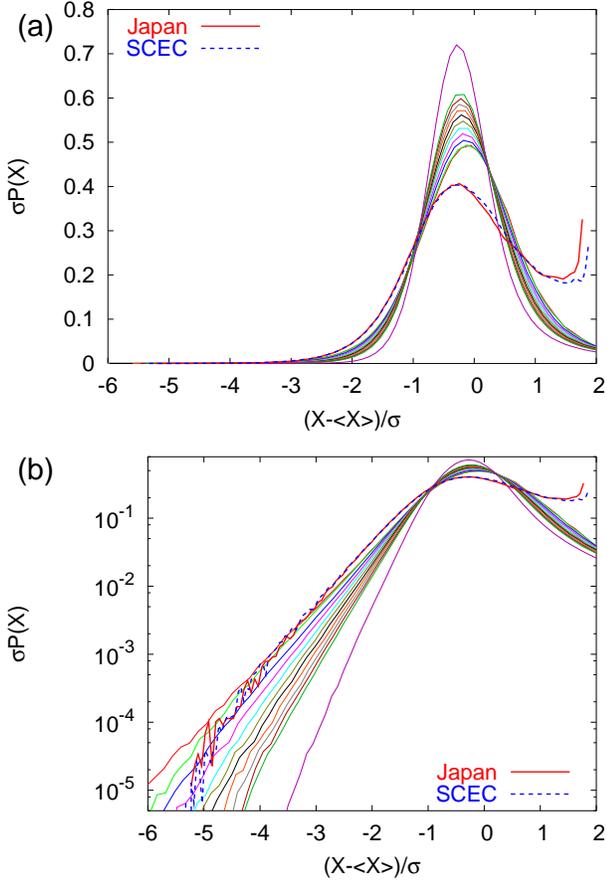}
\caption{\label{fgPoi} (Color online)  The linear-linear (a) and log-linear (b) plots
of $\sigma P(X)$ versus $(X-\langle X \rangle )/\sigma$, where $X$ stands for  $\Pi(\phi)$ for $\phi \approx 0$. The results for Japan (M$\geq 3.5$) and SCEC (M$\geq 2.0$) are plotted along 
with those deduced from a series of independent and identicaly distributed $(M_0)_k$ sampled 
from a Poisson distribution with mean value 5, 10, 20, 30, 40, 50, 60, 70, 80, 90, 100 and 200:
from the bottom to the top in the maxima appearing in (a), respectively, and 
from the upper to the lower left branch in (b), respectively.}
\end{figure}

\begin{figure}
\includegraphics{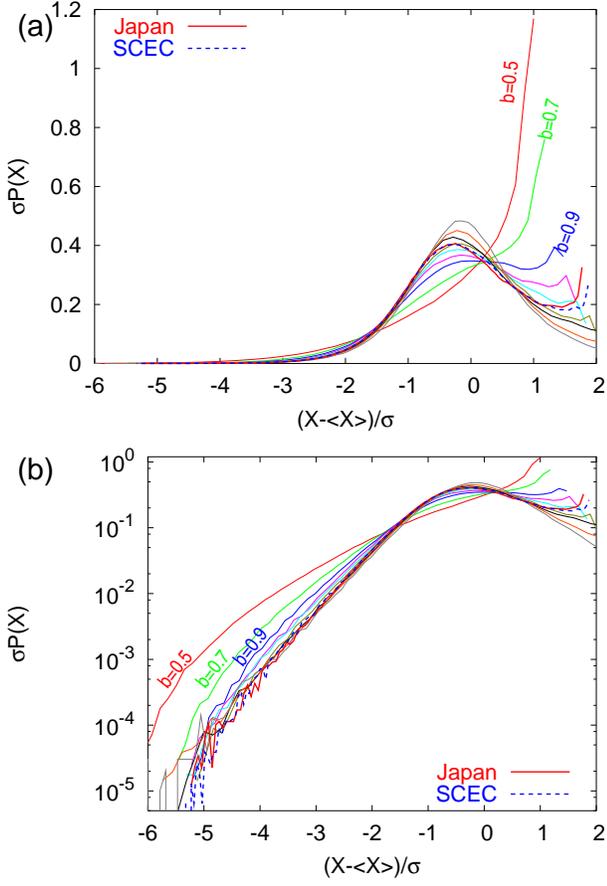}
\caption{\label{fgGuRi} (Color online)  The linear-linear (a) and log-linear (b) plots
of $\sigma P(X)$ versus $(X-\langle X \rangle )/\sigma$, where $X$ stands for  $\Pi(\phi)$ for $\phi \approx 0$. The results for Japan ($M\geq 3.5$) and SCEC ($M\geq 2.0$) are plotted along 
with those deduced from shuffled artificially generated EQ data obeying the Gutenberg-Richter law for 
various values of the exponent $b=$0.5, 0.7, 0.9, 1.0, 1.1, 1.2, 1.3, 1.5 and 2.0 from the lower to the 
upper curve at the value $(X-\langle X \rangle )/\sigma \approx -0.5$ in (a), and from the upper to the lower curve at the value $(X-\langle X \rangle )/\sigma \approx -4$ in (b).}
\end{figure}

\begin{figure}
\includegraphics{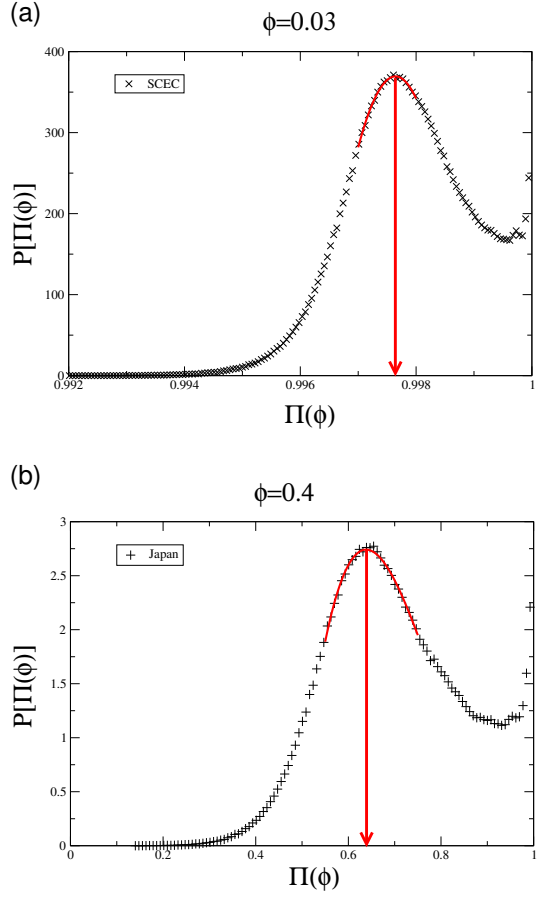}
\caption{\label{fg7} (Color online)  Two examples of the procedure used in the determination of  $\Pi_p(\phi)$.
(a) For SCEC at $\phi=0.03$, (b) For Japan at $\phi=0.4$. The (red) curves in each case, show the cubic polynomial fit which was used in the range $[a,b]$ around the maximum. The (red) arrow indicates the position of  $\Pi_p(\phi)$.}
\end{figure}

The  calculation of $\Pi(\phi)$ was made, as mentioned  in  Section IV, for an event taking time windows from 6 to  40
consecutive events and this process was performed for all
the  events  by  scanning the whole catalogue. This
procedure  resulted in the calculation, for  each  $\phi$-value
and  each  catalogue, of more than $10^6$  $\Pi(\phi)$-values,  whose
probability  density  function (PDF)  was  determined  by
using  the  computer code  {\tt histogram} of Ref.\cite{heg99} with  a
number  of  bins  proportional to $N^{1/3}$, where $N$  is  the
number  of  $\Pi(\phi)$-values  (cf. this point,  i.e.,  that  the
number  of  bins  should  be  proportional  to  $N^{1/3}$,  is
discussed in Ref. \cite{mer99}). This method resulted in  the  PDFs
shown in Fig.\ref{fg7}  as well as
in  those  depicted in Figs.\ref{fg2}(a),\ref{fg2}(b), \ref{fg4}, \ref{fg5}(b), and \ref{fg6}.
   Due  to  the
intrinsic fluctuations of the values of the PDF  (because
$N$  is  still finite), a direct determination of the value
 $\Pi_p(\phi)$ where the PDF maximizes, just by simply taking the
maximum  value  of  the calculated  PDF, may  lead  to
erroneous values of $\Pi_p(\phi)$. One should consider the  general
trend of the PDF as a whole, which can definitely show  a
more  accurate  and  stable  value  of  $\Pi_p(\phi)$.  Thus,   the
procedure we applied for the determination of $\Pi_p(\phi)$  was  as
follows:  For  each  $\phi$-value, a region  $[a,b]$  around  the
maximum was selected (examples are shown in Figs.\ref{fg7}(a)  and
(b))  and then a cubic polynomial, $p(x)=a+bx+cx^2+dx^3$, was
used  to fit the PDF values in this region. (Close enough
to  the maximum, a parabolic fit could be also good since
$f(x)=f_{max}-|f^{''}_{max} | (x-x_{max})^2/2$,  but  in  view  of  the  PDF
asymmetry the cubic polynomial used, provides a better
approximation in the whole region $[a,b]$.)  The  value  of
$\Pi_p(\phi)$ was determined through the direct maximization of this
cubic polynomial, i.e, $\Pi_p(\phi)=\frac{-2c-\sqrt{4c^2-12bd}}{6d}$. 
The values of  $\Pi_p(\phi)$ shown
in  Fig.\ref{fg2}(c) have been obtained by means
of  such a procedure. Finally, we note that, due  to  the
fitting procedure involved and the relative arbitrariness
in  the definition of [a,b], the estimation error of  $\Pi_p(\phi)$
is  more  or less comparable to its percentage  deviation
from  $\Pi_{th}(\phi)$, depicted in Fig.\ref{fg2}(d).  Thus,
we  can  state  that $\Pi_p(\phi)$  and  $\Pi_{th}(\phi)$  are  experimentally
indistinguishable, which strengthens the  statement  that
Eq.(2) -which  has  been  used  for
estimating $\Pi_{th}(\phi)$- holds for EQs.

\end{document}